# A Cascade Transformer-based Model for 3D Dose Distribution Prediction in Head and Neck Cancer Radiotherapy

Tara Gheshlaghi[1], Shahabedin Nabavi[1], Samire Shirzadikia[2], Mohsen Ebrahimi Moghaddam[1], Nima Rostampour[2]

1. Faculty of Computer Science and Engineering, Shahid Beheshti University, Tehran, Iran.
2. Department of Medical Physics, School of Medicine, Kermanshah University of Medical Sciences, Kermanshah, Iran.

**Corresponding Author:** Nima Rostampour (PhD)

**Address:** Department of Medical Physics, School of Medicine, Kermanshah University of Medical Sciences, Kermanshah, Iran.

**Email:** nima.rostampour@kums.ac.ir

**Phone:** +98 918 333 2075

**Abstract**

Background: Radiation therapy is the primary method used to treat cancer in the clinic. Its goal is to deliver a precise dose to the planning target volume (PTV) while protecting the surrounding organs at risk (OARs). However, the traditional workflow used by dosimetrists to plan the treatment is time-consuming and subjective, requiring iterative adjustments based on their experience. Deep learning methods can be used to predict dose distribution maps to address these limitations.

Methods: The study proposes a cascade model for organs at risk segmentation and dose distribution prediction. An encoder–decoder network has been developed for the segmentation task, in which the encoder consists of transformer blocks, and the decoder uses multi–scale convolutional blocks. Another cascade encoder–decoder network has been proposed for dose distribution prediction using a pyramid architecture. The proposed model has been evaluated using an in–house head and neck cancer dataset of 96 patients and OpenKBP, a public head and neck cancer dataset of 340 patients.

Results: The segmentation subnet achieved 0.79 and 2.71 for Dice and HD95 scores, respectively. This subnet outperformed the existing baselines. The dose distribution prediction subnet outperformed the winner of the OpenKBP2020 competition with 2.77 and 1.79 for dose and DVH scores, respectively.

Conclusions: The predicted dose maps showed good coincidence with ground-truth, with a superiority after linking with the auxiliary segmentation task. The proposed model outperformed state-of-the-art methods, especially in regions with low prescribed doses.

The codes are available at https://github.com/GhTara/Dose_Prediction.

**Keywords:** Radiation therapy, Dose distribution prediction, Deep learning, Transformer neural network, Multi-scale convolutional neural network, Head and neck cancer

## 1. Introduction

Radiation therapy (RT) has been nominated as one of the primary treatment modalities for patients with malignant tumors. The development of novel RT techniques such as intensity-modulated radiation therapy (IMRT) has significantly improved the treatment plan quality with high homogeneous dose distribution and spatial precision over the last few years (Osman and Tamam 2022). With these advancements, desirable therapeutic effects and local tumor control have been achieved by delivering a high dose to the planning target volume (PTV) with superior organs at risk (OARs) sparing (Webb 1991, Otto 2008). However, the collaboration of physicians and oncologists and repeatedly adjusting the beam parameters are crucial to the balance between tumor control and normal organ toxicities for planning a treatment strategy (Zhan, Xiao et al. 2022). Moreover, the skill and knowledge of planners touch the plan quality, resulting in a significant inconsistency among inter-observers to derive an optimal clinical treatment plan (Geets, Daisne et al. 2005, Brouwer, Steenbakkers et al. 2012). In particular, manual volume delineation is highly time-consuming (Vorwerk, Zink et al. 2014) and laborious in cancers with a complex anatomical structure such as head and neck (H&N) (Van Dijk, Van den Bosch et al. 2020), due to a sizable PTV and numerous major OARs nearby it (Paulino, Koshy et al. 2005, Rancati, Schwarz et al. 2010). These issues extremely prolonged the required time for treatment planning, which not only deters the employment of adaptive radiation therapy (ART) (Yan, Vicini et al. 1997), but also delays treatment delivery. Both of these outcomes are likely to have negative effects on tumor control as well as the patient's quality of life (Yang, Hu et al. 2013, Ferreira, Olasolo et al. 2015). Another downside is depending on the current treatment planning systems on dose-volume histogram (DVH) which can't portray the heterogeneity of the spatial dose distribution inside a structure and ignores the non-contoured structures (Mashayekhi, Tapia et al. 2022). Researchers have been working on these issues for many years. Recently, new methods known as knowledge-based planning (KBP) have been introduced to automatically dose predict and improve the quality of planning. These methods primarily develop a new treatment plan based on previous patients' data with similar anatomical features. A typical clinical assessment index used in KBP approaches is the DVH (Cotrutz and Xing 2002, Carrasco, Jornet et al. 2012). Although some new handicraft features such as distance-to-target histograms (DTH) and overlap volume histograms (OVH) were employed to better match a new patient with previous ones and drive its dose distribution (Thomas-Kaskel, Zeiser et al. 2006, Wu, Mcnutt et al. 2011, Yuan, Ge et al. 2012), the KBP methods have several restrictions. First, the size and diversity of historical patients determined the performance of KBP-based methods. Second, these methods are extremely based on manually derived characteristics from the dose distribution patterns, extracting these features takes a lot of effort and isn't very accurate. As another issue, the knowledge-based approaches derive one- or two-dimensional features, such as DVH and OVH (Mahmood, Babier et al. 2018) which are blind to spatial data of the dose distribution. This is because the three-dimensional dosimetric features are compressed and asymmetrically mapped, leading to a loss of information. Moreover, the details of the anatomical structure and position surrounding the tumor were overlooked, ignoring the impact of neighboring organs.

Deep learning (DL) techniques have made substantial advancements in medical image processing. Convolutional neural networks (CNN) and their variations are a prominent development of deep learning (LeCun, Boser et al. 1989). Many studies have used CNN algorithms for medical image segmentation and prediction of radiation dose in H&N cancer (Ronneberger, Fischer et al. 2015, Chen, Men et al. 2019, Fan, Wang et al. 2019, Hatamizadeh, Nath et al. 2022). The encoder-decoder architecture serves as the backbone for many of these techniques. For example, a CNN-based network called U-Net, which was first developed for segmentation, was modified for dose distribution prediction by adding skip-connections between encoder-decoder structures (Ronneberger, Fischer et al. 2015). This simple but effective network has greatly gained the researchers' attention as far as many studies are based on various modified U-Net to predict the optimal dose distribution and PTV and OARs masks (Kearney, Chan et al. 2018, Nguyen, Long et al. 2019, Shao, Zhang et al. 2020).

Despite the reduced dependency of DL-based algorithms on the manually derived characteristics and fulfillment of a certain level of automated process of radiation planning, some restrictions must be taken into consideration. For instance, some extra anatomical features with CT images must be supplied into the network to guarantee accurate prediction of dose distribution, which means radiotherapists' knowledge is still needed for dose prediction (Li, Peng et al. 2022). As the most important ones, the previous works evaluated segmentation and the predicted dose distribution separately, ignoring the relationship between them. The link between dose distribution and segmentation necessitates a clinically optimal dose distribution, considering that the amount of received dose by the PTV is highly dependent on its feature (such as shape and size) as well as the relative location of OARs to reach the main goal of radiotherapy, which is delivering the highest fraction of the prescribed dose to the tumor while the least one to OARs. It is necessary to precisely segment the PTV and OARs to predict the dose distribution accurately, and the dose distribution map anatomically portrays the structures' outlines, showing a strong correlation between these two tasks (Li, Peng et al. 2022).

We present two models to delineate OARs masks and predict dose distribution automatically. Moreover, we developed a linked model to evaluate how anatomical information obtained from the segmentation task can impact the dose prediction. The contributions of this study are as follows:

(1) We proposed a model consisting of multi-scale convolutional units and transformer blocks, which can simultaneously extract local and global features. The simultaneous extraction of local and global features is one of the challenges in deep learning that we tried to address using this approach. This approach was used and evaluated for automatic OARs segmentation.
(2) We proposed a cascade encoder-decoder model, including transformer blocks and multi-scale convolutional units based on a pyramid structure to predict dose distribution. The extensive experiments show that the proposed method outperforms the winner of the OpenKBP2020 challenge and other previous studies.
(3) In designing the proposed model, we tried to automate the dose distribution prediction as much as possible. Therefore, the model can automatically segment OARs by receiving CT images and corresponding PTVs of the patient and use the extracted OARs to predict dose distribution.

**2. Materials and Methods**

Our proposed cascade architecture consists of two sub-networks of OARs segmentation and dose distribution prediction. The overview of the proposed model is shown in Figure 1. In this section, the problem is formulated first. Then the details of the model and its training process are described.

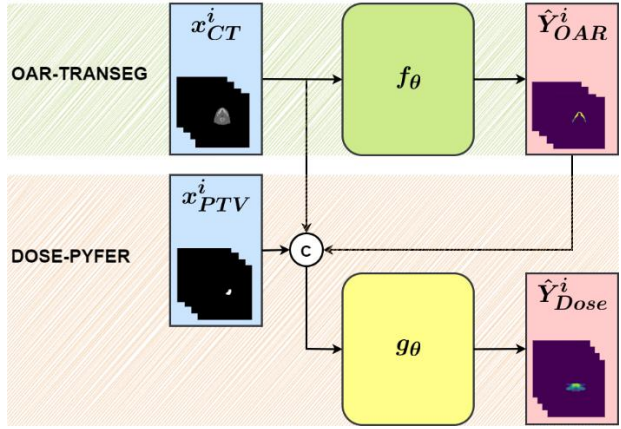

**Fig. 1.** The overview of the proposed model.

*2.1. Problem Formulation*

The subnet related to OARs segmentation can segment seven OARs by receiving a volume of CT images of a patient. The task of the OARs segmentation subnet is defined as follows:

$$f_\theta: x_{CT}^i \rightarrow \hat{Y}_{OAR}^{ij}$$

where $x_{CT}^i$ shows the $i^{th}$ input CT volume, and $\hat{Y}_{OAR}^{ij}$ represents the OARs extracted by this subnet for the $i^{th}$ input. The value of $j$ can also be 1 to 8, which indicates the corresponding seven different OARs and the background of images.

The task of the dose distribution prediction subnet is defined as follows:

$$g_\theta: Concat(x_{CT}^i, \hat{Y}_{OAR}^{ij}, x_{PTV}^i) \rightarrow \hat{Y}_{Dose}^i$$

where the input of this subnet is the concatenated of $i^{th}$ CT volume, the output of the segmentation subnet ($\hat{Y}_{OAR}^{ij}$) and contoured PTVs ($x_{PTV}^i$), and the output of this subnet is the corresponding predicted dose distribution map ($\hat{Y}_{Dose}^i$).

*2.2. Proposed Model Architecture*

*2.2.1. OARs Segmentation Subnet*

An encoder-decoder architecture is leveraged for segmenting seven OARs, including the brainstem, spinal cord, right and left parotid, esophagus, larynx and mandible. This subnet uses a hybrid architecture, including the vision transformers (Dosovitskiy, Beyer et al. 2020) for the encoder and a decoder based on multiscale convolutions. The architecture of this subnet, which we named OAR-TRANSEG, is shown in Figure 2.

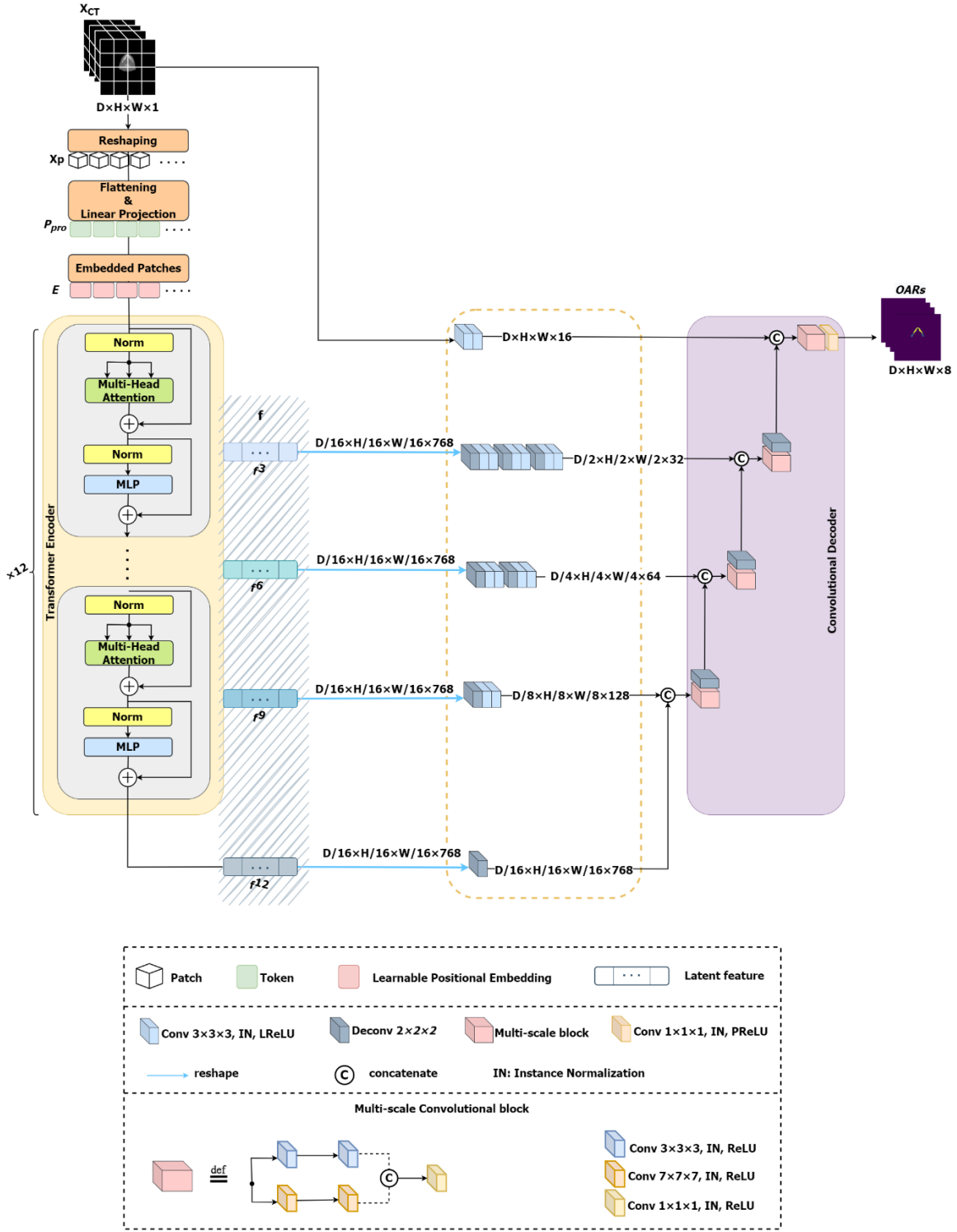

**Fig. 2.** Architecture of the OARs segmentation subnet.

This subnet generates $\hat{Y}_{OAR}^{ij}$, which contains eight volumes corresponding to seven OARs and the background of images as output by receiving $x_{CT}^i \in \mathbb{R}^{H \times W \times D \times C}$ as input. $(H, W, D, C) = (128, 128, 128, 1)$ where $H$, $W$, $D$ and $C$ represent height, width, depth (volume resolution) and number of channels/volumes respectively. The input volume is first divided into non-overlapping patches of size $(P, P, P) = (16, 16, 16)$. Thus, the input volume $x_{CT}^i$ is converted into a sequence of patches in the form of $x_P^{ij}$, where the dimension of all patches together is $\mathbb{R}^{N \times (P^3 \cdot C)}$ for $N = \frac{H \times W \times D}{P^3}$. Then, after flattening, patches are reduced in dimensions by passing through a linear layer with $k = 768$ neurons to a projected embedding $P_{pro}^i \in \mathbb{R}^{N \times K}$. To preserve the positional information of each $j^{th}$ patch, the 1D trainable embedding $E^i \in \mathbb{R}^{N \times K}$ is added to the projected embedding $P_{pro}^i$ as follows:

$$P_E^{ij} = P_{pro}^{ij} + E^{ij} \qquad (1)$$

The $P_E^i$ obtained after passing through $l = 12$ layers of transformers leads to the production of latent vectors $f^l$, which are the output of the $l^{th}$ layer of the transformer. Each transformer layer includes multi-head self-attention (MSA) sublayers with 12 heads and multilayer perceptron. MSA sublayers extract global information from patches by comparing pairs of $P_E^{ij}$.

The connection between the encoder and decoder is established through several skip connections that combine the latent features of $f^3$, $f^6$, $f^9$ and $f^{12}$ with the corresponding feature maps of the decoder. To convert the resolution of latent features of the encoder to the appropriate resolution of the feature map of the decoder, after reshaping $f^3$, $f^6$, $f^9$ and $f^{12}$ from $\frac{HWD}{P^3} \times K$ to $\frac{D}{P} \times \frac{H}{P} \times \frac{W}{P} \times K$, we have used $2 \times 2 \times 2$ deconvolutional and $3 \times 3 \times 3$ convolutional layers (See Figure 2). The combined feature maps are entered into a multiscale convolutional block and subjected to convolution operation with kernels $3 \times 3 \times 3$ and $7 \times 7 \times 7$. The reason for using multiscale convolution is to create a variety of features from the point of view of producing local and more global features. Finally, $\hat{Y}_{OAR}^{ij} \in \mathbb{R}^{H \times W \times D \times 8}$ includes eight volumes for seven OARs and the background of images.

*2.2.3. Dose Distribution Prediction Subnet*

A cascade architecture consisting of two components is used for the dose distribution prediction task. The first component is a pre-trained U-Net network (Liu, Zhang et al. 2021), which is used for feature extraction, and its weights are frozen during the training process. The second component is the main part of dose distribution prediction subnet, which has an encoder-decoder architecture. The architecture of this component is similar to the OAR-TRANSEG with a difference. We used a pyramid-based structure in the decoder section of this component to improve efficiency. A view of the dose distribution prediction subnet, which we called DOSE-PYFER, is shown in Figure 3.

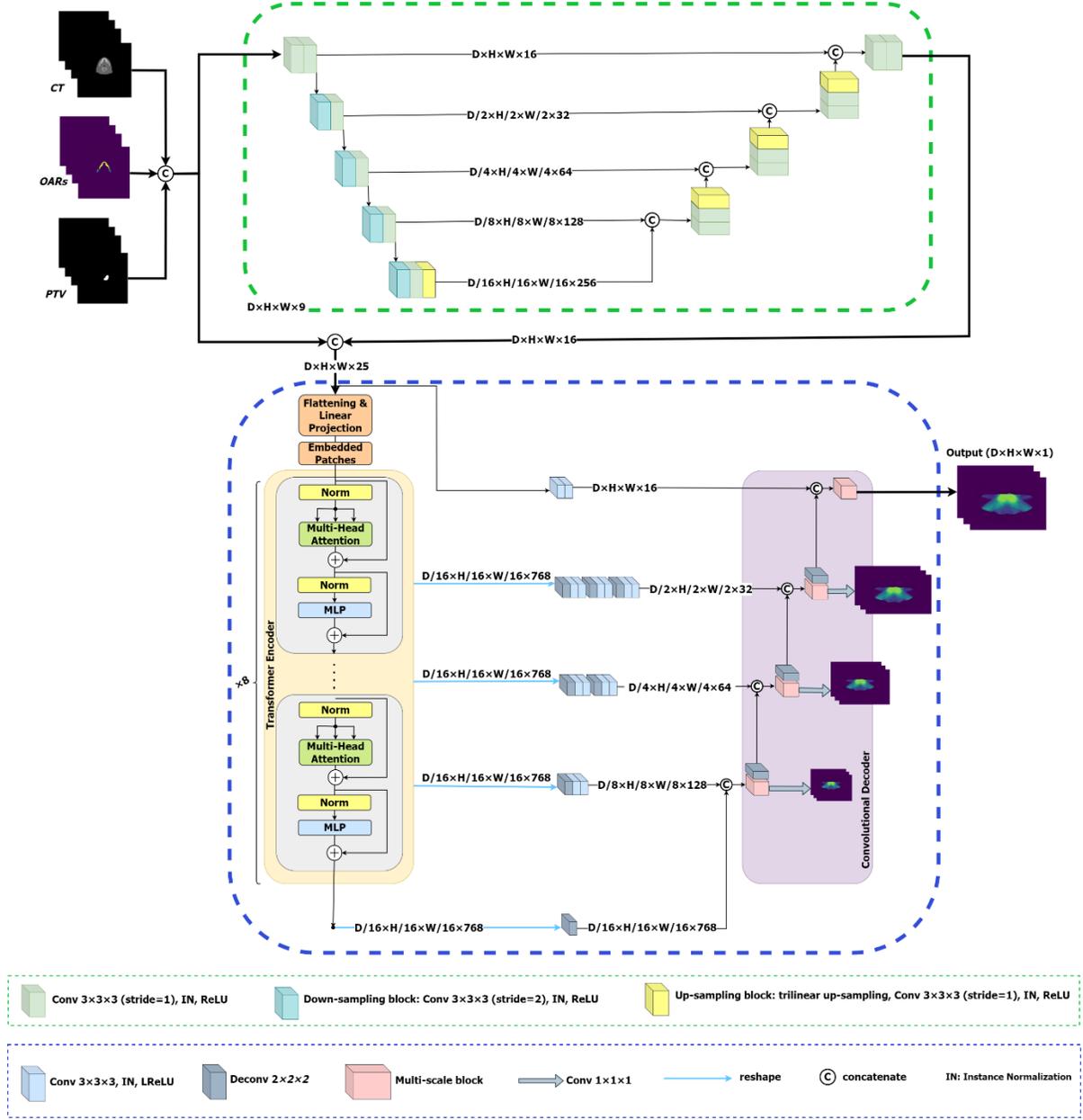

**Fig. 3.** Architecture of the dose distribution prediction subnet.

The input of this subnet is $x_{COP}^i \in \mathbb{R}^{H \times W \times D \times C}$, which is obtained as follows:

$$x_{COP}^i = Concat(x_{CT}^i, \hat{Y}_{OAR}^{ij}, x_{PTV}^i) \qquad (2)$$

where $(H, W, D) = (128, 128, 128)$ and $C = 9$ include seven volumes of OARs, one PTV and one CT volume.

After feeding $x_{COP}^i$ to the first component, the generated output is concatenated with $x_{COP}^i$ to provide a more information-rich input to the second component. In the second component, eight transformer layers with six heads are used to produce $f^l$ as latent features. Then the latent features $f^2$, $f^4$, $f^6$ and $f^8$ are connected to the feature maps of the corresponding layer in the decoder through skip connections.

We considered the pyramid-like structure for the second component's decoder to solve the training instability problem. This problem occurs when small changes in the training data cause drastic changes in the cost function and the final prediction. Intermediate outputs of different decoder levels with lower resolution than the final output or the final dose distribution map are used to solve this problem. In other words, we involve intermediate gradients to generate intermediate outputs, which makes the model training process more stable, and the outputs of the intermediate layers are encouraged to resemble the ground truth with their corresponding scale. Trilinear interpolation is used to have the ground truth in the pyramid structure or smaller resolutions because the dose distribution maps are volumetric.

*2.3. Objective Functions and Training Details*

For OAR-TRANSEG subnet, the objective functions $\mathcal{L}_{OAR}$ is defined as follows:

$$\mathcal{L}_{oar}(Y_{OAR}^i, \hat{Y}_{OAR}^i) = \\ = 1 - \frac{2}{J}\sum_{j=1}^{J}\frac{\sum_{k=1}^{K}Y_{OAR}^{ij}(k)\cdot\hat{Y}_{OAR}^{ij}(k)}{\sum_{k=1}^{K}(Y_{OAR}^{ij}(k))^2 + \sum_{k=1}^{K}(\hat{Y}_{OAR}^{ij}(k))^2} \\ - \frac{1}{K}\sum_{k=1}^{K}\sum_{j=1}^{J}Y_{OAR}^{ij}(k)\cdot log(\hat{Y}_{OAR}^{ij}(k)) \tag{3}$$

where $Y_{OAR}^i$ and $\hat{Y}_{OAR}^i$ are the $i^{th}$ ground-truth and the output of OAR-TRANSEG subnet for the $k^{th}$ voxel of the $j^{th}$ OARs, respectively. Besides, $J$ and $K$ indicate the number of OARs and the number of voxels, respectively.

For DOSE-PYFER subnet, the objective function $\mathcal{L}_{Dose}$ is considered as follows:

$$\mathcal{L}_{dose} = \lambda_1 \mathcal{L}_{out} + \lambda_2 \mathcal{L}_{ds} \qquad (4)$$

where $\mathcal{L}_{ds}$ (Equation 5) indicates the loss function in different scales, and $\mathcal{L}_{out}$ (Equation 6) is the loss function related to the final output. Based on ablation studies, the values of 8 and 10 were considered for $\lambda_1$ and $\lambda_2$ coefficients, respectively.

$$\mathcal{L}_{ds} = \frac{\sum_{s=1}^{N-1} \left\| Y_{Dose}^i(s) - \hat{Y}_{Dose}^i(s) \right\|_1}{N-1} \qquad (5)$$

$$\mathcal{L}_{out} = \left\| Y_{Dose}^i(N) - \hat{Y}_{Dose}^i(N) \right\|_1 \qquad (6)$$

where considering $N$ outputs in the different scales, $Y_{Dose}^i(s)$ and $\hat{Y}_{Dose}^i(s)$ show the ground-truth and predicted dose distribution, respectively, in the $s^{th}$ scale for $i^{th}$ input.

The proposed model was implemented in Python programming language using PyTorch (Paszke, Gross et al. 2019), PyTorch Lightning, and MONAI (Cardoso, Li et al. 2022) libraries. Ray library (Moritz, Nishihara et al. 2018) was used to obtain hyperparameters, and MLflow library (Zaharia, Chen et al. 2018) and Databricks cloud environment were leveraged to do experiments and compare results. The models were trained in the Google Colab Pro environment with GPU and 32 GB of memory. The model training in different experiments has been done with a maximum batch size of 2 due to hardware limitations. Besides, AdamW (Loshchilov and Hutter 2017) and Adam8bit (Dettmers, Lewis et al. 2021) optimizers were used in training OAR-TRANSEG and DOSE-PYFER subnets for 200 epochs. Learning rate (lr) and weight decay (wd) were considered $lr = 10^{-4}$ and $wd = 10^{-5}$ for OAR-TRANSEG, and $lr = 6.131 \times 10^{-4}$ and $wd = 1.63 \times 10^{-4}$ for DOSE-PYFER. During the training of the OAR-TRANSEG subnet and in some experiments related to the DOSE-PYFER subnet, data augmentation methods such as the random shift of voxel intensity, the random crop of input volumes, the random flip and 90-degree rotations were used.

*2.4. Datasets*

Our experiments were mainly two tasks of OARs segmentation and prediction of dose distribution which were conducted on two datasets of in-house radiation therapy data of 96 H&N patients with the original CT image, PTV, and OARs masks contoured by experienced oncologists, and their corresponding dose distribution maps, and a public *OpenKBP-2020AAPM Grand Challenge*[1] dataset (Babier, Zhang et al. 2021) consisting 340 subjects for H&N cancers with contoured CT images and their dose distribution maps. The OAR segmentation experiments were performed on both mentioned datasets. In the case of the in-house dataset, we used the zero padding strategy to set the z-dimension of all volumes to 128 and the data augmentation strategy to deal with the small size of the in-house dataset. Eighty subjects from the in-house dataset were considered the training set, and the rest were used for the test set. For the dose distribution prediction, we only used the OpenKBP-2020AAPM Grand Challenge dataset with the same presented splits of training, validation and testing sets.

*2.5. Evaluation metrics*

For OARs segmentation, the performance of our model was evaluated with the clinically conventional metrics as follows:

(1) Dice metric displays the regional similarities between the segmented images and the ground-truth images.

$$Dice(G, P) = \frac{2 \sum_{i=1}^{I} G_i P_i}{\sum_{i=1}^{I} G_i^2 + \sum_{i=1}^{I} P_i^2} \qquad (7)$$

where $P_i$ refers to the predicted voxel and $G_i$ refers to the ground-truth voxel. The Dice is a value between 0 and 1, with a higher value indicating better performance of the model.

(2) *95% Hausdorff distance (HD95)* is a measure for determining the dissimilarity surface between the predicted labels and true labels.

$$HD95(G, P) = max \left\{ \max_{g \in G} \min_{p \in P} \|g - p\|, \max_{p \in P} \min_{g \in G} \|p - g\| \right\} \qquad (8)$$

The lower HD95 indicates a better segmentation result.

The performance of the dose prediction subnet was evaluated based on two main criteria. The dose score metric measures the mean absolute error between the predicted dose distribution and the corresponding ground-truth values. The DVH score calculates the mean absolute error of five DVH metrics ($D_{0.1cc}$, $D_{mean}$, $D_{1\%}$, $D_{95\%}$, and $D_{99\%}$). The model that achieves the lowest dose and DVH scores is considered to have the best performance. Besides, the differences between the five DVH parameters were computed and reported separately. $D_{0.1cc}$ and $D_{mean}$ were calculated for the OARs, and $D_{1\%}$, $D_{95\%}$, and $D_{99\%}$ were measured for the PTVs. The dose score and DVH score are defined as follows:

$$\text{Dose Score} = \sum_{p=1}^{N} \frac{\frac{\sum_{i=1}^{I} |G_{p,i} - P_{p,i}|}{I}}{N} \tag{9}$$

$$\text{DVH Score} = \frac{1}{\sum_{p=1}^{N} \sum_{v \in V_{p,OAR} \cup V_{p,PTV}} |C_v|} \sum_{p=1}^{N} \sum_{v \in V_{p,OAR} \cup V_{p,PTV}} \sum_{c \in C_v} |D_c^v(G_p) - D_c^v(P_p)|, \forall c \in C_v, \forall v \in V_{p,OAR} \cup V_{p,PTV} \tag{10}$$

---
[1] OpenKBP-2020 AAPM Grand Challenge dataset is available at https://www.aapm.org/GrandChallenge/OpenKBP/

Besides, we used a paired t-test to compute the p-value for statistical comparison of our predicted results with the previous results and the ground truth.

## 3. Results

### 3.1 Evaluation of OARs Segmentation

#### 3.1.1 Ablation Study

To achieve the best architecture, we performed an ablation study in the following settings:

- Transformer-based encoder + convolutional decoder
- Transformer-based encoder + multi-scale convolutional decoder
- Transformer-based encoder + multi-scale convolutions at skip-connections

We evaluated the proposed model using data from 30 subjects randomly selected from the dataset, including 24 subjects for the training and six for the testing. Table 1 presents the obtained results. These experiments examined the impact of including multi-scale convolutional units in the model architecture. The transformer-based encoder + multi-scale convolutional decoder was selected as the optimal model (OAR-TRANSEG) due to the better value of the Dice metric.

**Table 1**. The quantitative results of ablation study for OAR-TRANSEG subnet. The number of model parameters is in millions (M).

| Method | Dice | Training loss function | Model Parameters (M) |
|---|---|---|---|
| Transformer-based encoder + convolutional decoder | 0.360 | 0.903 | 92.8 |
| Transformer-based encoder + multi-scale convolutional decoder | **0.707** | **0.384** | 115 |
| Transformer-based encoder + multi-scale convolutions at skip-connections | 0.667 | 0.483 | 147 |

*3.1.2 Results of OAR-TRANSEG*

The results of the proposed OARs segmentation model are shown in Table 2. These results are reported separately for seven OARs based on the Dice and HD95 metrics. The OAR-TRANSEG subnet was also evaluated using transfer learning. In the experiment with transfer learning, the network was trained on the in-house dataset, and then the network weights were stored. In the next step, training and testing was done on the OpenKBP dataset, considering the stored weights as initial weights. The results of this experiment are also listed in Table 2. Transfer learning could improve the value obtained for the Dice metric in the overall mode. Examples of the results obtained from the OAR-TRANSEG subnet versus grand-truth are shown in Figure 4.

**Table 2**. Quantitative comparison of the proposed model for OARs segmentation in terms of Dice and HD95. The results with and without transfer learning on the in-house dataset are listed.

| OAR | Method | | | |
|---|---|---|---|---|
| | OAR-TRANSEG (Without Transfer Learning) | | OAR-TRANSEG (With Transfer Learning) | |
| | HD95 | Dice | HD95 | Dice |
| Brainstem | **2.3391** | **0.7744** | 2.3717 | 0.7682 |
| Spinal Cord | **3.9108** | **0.7631** | 4.2138 | 0.7621 |
| Right Parotid | 2.7243 | 0.7683 | **2.3086** | **0.7741** |
| Left Parotid | 3.4235 | **0.7613** | **2.6795** | 0.7641 |
| Esophagus | **5.8140** | **0.6152** | 7.6695 | 0.6155 |
| Larynx | **4.5748** | **0.6247** | 5.0864 | 0.6204 |
| Mandible | **1.9029** | 0.8767 | 1.9971 | 0.8815 |
| Overall | **3.5271** | 0.7405 | 3.7609 | **0.7408** |

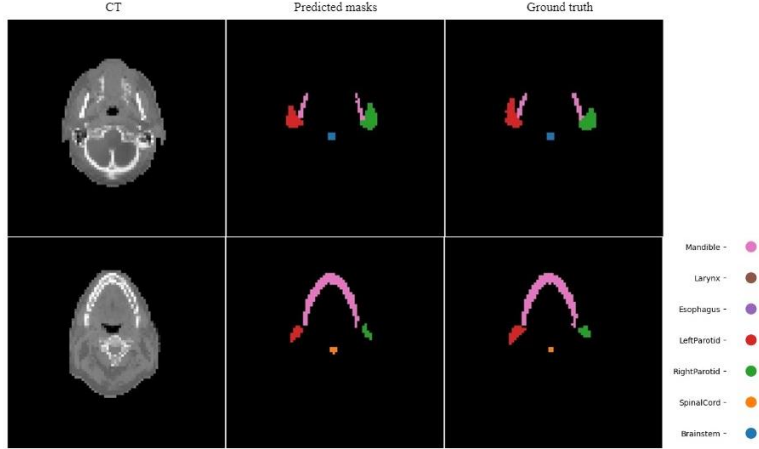

**Fig.4.** Examples of the results obtained from the OAR-TRANSEG subnet versus grand-truth.

*3.1.3 The proposed OAR-TRANSEG vs. State-of-the-art Method*

We compared our proposed model (OAR-TRANSEG) with the state-of-the-art model (3D ResU-Net) (Isler, Lisle et al. 2022), which is a convolutional network with residual blocks evaluated on the OpenKBP dataset, to show the superiority of our model in the OARs segmentation task. Since the 3D ResU-Net model considered only the five OARs from the OpenKBP dataset, we conducted a comparison of these five OARs for fairness. As shown in Table 3, the proposed model can segment the OARs more accurately than the 3D ResU-Net model. The proposed model achieved a 20% improvement in the HD95 value, with the most significant improvement seen in the brainstem and spinal cord.

**Table 3**. Comparison of the proposed model with the state-of-the-art model for OARs segmentation task.

| OAR | Method | | | |
|---|---|---|---|---|
| | Proposed OAR-TRANSEG Model | | 3D ResU-Net Model | |
| | Dice | HD95 | Dice | HD95 |
| Brainstem | 0.7682 | **2.3717** | **0.80** | 3.94 |
| Spinal Cord | **0.7621** | **4.2138** | 0.75 | 5.97 |
| Right Parotid | **0.7741** | **2.3086** | 0.76 | 2.31 |
| Left Parotid | **0.7641** | 2.6795 | 0.75 | **2.28** |
| Mandible | **0.8815** | 1.9971 | 0.86 | **1.78** |
| Total | **0.7900** | **2.7141** | 0.78 | 3.26 |

*3.2 Evaluation of the Dose Distribution Prediction*

*3.2.1 Ablation Study*

We investigated the effectiveness of several parameters, such as data augmentation, activation function, loss function, and cascade architecture, in the ablation studies regarding DOSE-PYFER. We evaluated the proposed model using data from 120 subjects randomly selected from the OpenKBP dataset, including 100 subjects for the training and 20 for the testing. The results of these tests are given in Table 4. The DVH curves produced by the model with different parameters are shown in Figure 5. These curves present a good coincidence of the dashed line as the predicted DVH with the solid line as the ground-truth in the proposed model. Moreover, a visual comparison of ablation experiments was displayed in Figure 6. The model predicted the dose maps showing minimal disparity from the ground-truth in both the high-dose PTV regions and low-dose OAR areas.

Table 4. Dose and DVH Scores in different ablation tests using OpenKBP. The check mark shows including the parameter in the experiment. The best results are highlighted in bold font.

|  | Huber LF | DA | L1 LF | Mish AF | Cascade Architecture | Increasing $\lambda_2$ | Dose Score | Dose DVH |
|---|---|---|---|---|---|---|---|---|
| DOSE_PYFER | ✓ |  |  |  |  |  | 5.69 | 3.71 |
|  | ✓ | ✓ |  |  |  |  | 5.48 | 4.23 |
|  |  | ✓ | ✓ |  |  |  | 4.98 | 2.93 |
|  |  |  | ✓ | ✓ | ✓ |  | 2.94 | 1.90 |
|  |  |  | ✓ | ✓ | ✓ | ✓ | **2.91** | **1.74** |

LF: Loss Function; DA: Data Augmentation; AF: Activation Function;

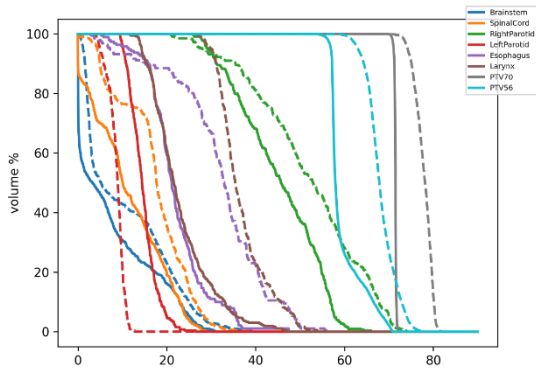

(a)

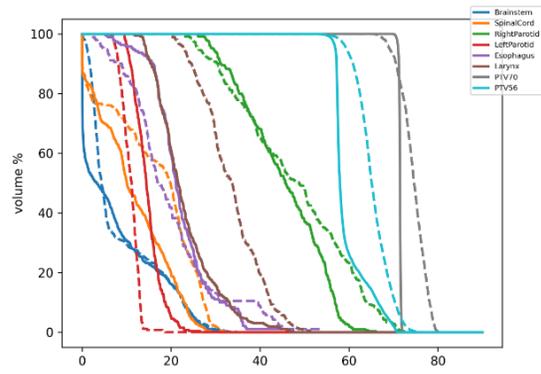

(b)

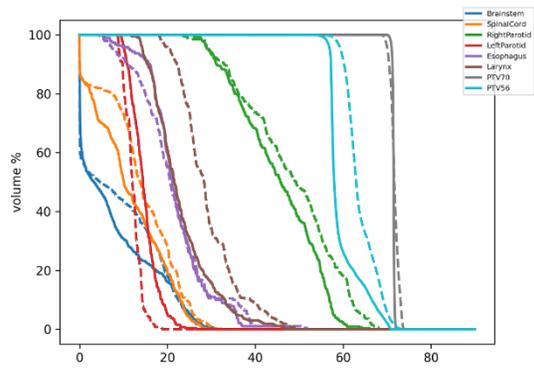

(c)

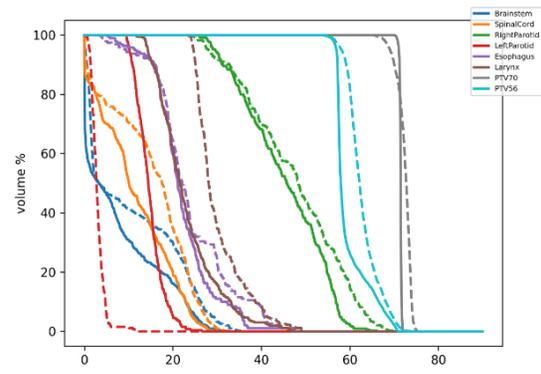

(d)

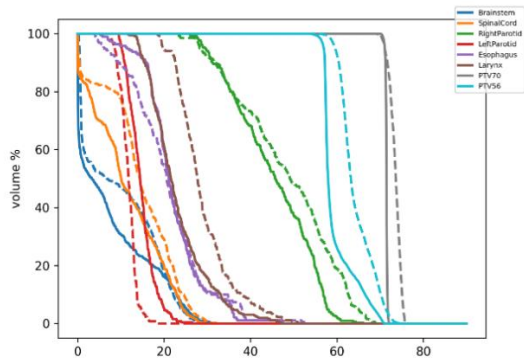

(e)

**Fig. 5.** DVH curve comparison of the model in different ablation tests.

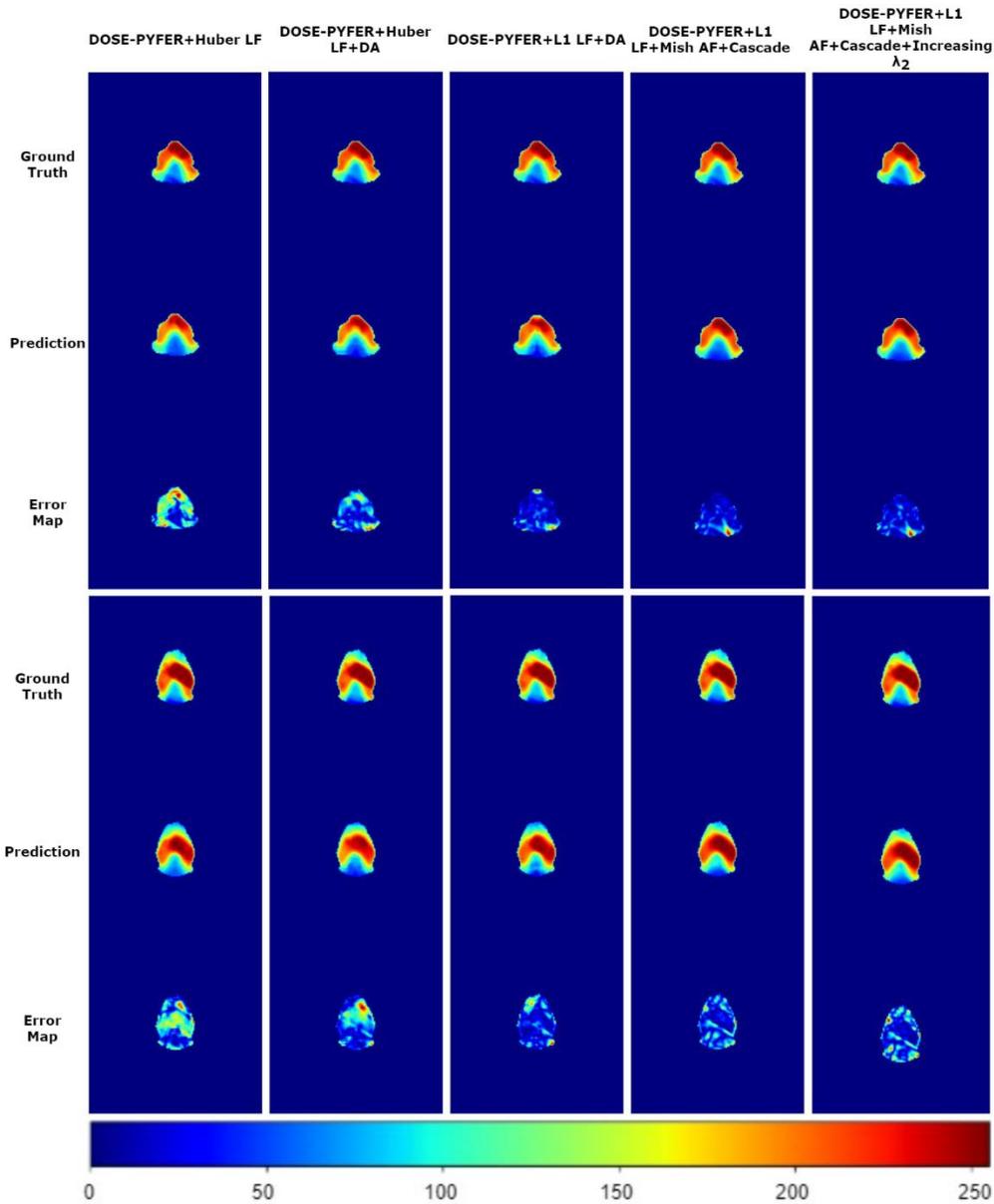

**Fig. 6.** The predicted dose map comparison of the model in different ablation tests.

### 3.2.2 The proposed DOSE-PYFER vs. GANs

We also developed architectures based on generative adversarial networks (GAN) to compare with the proposed architecture. We implemented a network based on convolutional networks with random weights for the generator part, similar to the MSG-GAN model (Karnewar and Wang 2020). We used the pre-trained ResNet architecture as the discriminator. In this way, we considered the outputs of the generator (four-scale outputs) as the inputs of the discriminating network in their corresponding scale. Besides, we used self-attention blocks in the discriminator similar to the DoseGAN architecture (Karnewar and Wang 2020). Finally, combining all these features, we obtained the best results for the architecture based on GANs, as shown in **Table 5**. The encoder-decoder-based networks showed better performance than GANs on the OpenKBP dataset (Jha, Sajeev et al. 2022).

**Table 5.** Dose Score obtained by GAN, conditional GAN (cGAN) and proposed architectures for OpenKBP dataset.

| Method | Dose score |
|---|---|
| GAN = The proposed DOSE-PYEFR as Generator + pre-trained ResNet as Discriminator | 6.579 |
| cGAN = The proposed DOSE-PYFER as Generator + pre-trained ResNet as Discriminator + considering input of generator as additional input for discriminator + self-attention gate | 3.317 |
| The Proposed DOSE-PYFER | 2.778 |
| The Proposed End-to-End Model (OAR-TRANSEG + DOSE-PYFER) | **2.643** |

*3.2.3 The proposed DOSE-PYFER vs. State-of-the-art Method*

The performance of the DOSE-PYFER for the dose distribution prediction was compared with the C3D model (Liu, Zhang et al. 2021), which presented the best performance among the state-of-the-art methods on the OpenKBP dataset. The C3D model measured a score of 2.86 for dose and 1.83 for DVH scores, whereas the DOSE-PYEFR could increase the performance with a dose score of 2.78 and a DVH score of 1.79. Besides, the DOSE-PYFER was a better dose predictor than the C3D model in the areas with low prescription doses. As indicated by the red arrow in Figure **7**, a higher dose was predicted using the C3D model, while the proposed model could calculate a dose closer to the ground truth. Similarly, the proposed method outperformed the prediction of dose maps for PTV. In this way, the data distribution was statistically compared using violin plots. The violin plots in Figure **8** compared the metrics of $D_{99}^v$ and $D_{95}^v$ as criteria for the minimum prescription dose in PTV. The proposed method showed the data distribution closer to zero on the violin plot than the C3D model, indicating a better prediction for the low-dose areas. The DVH curves in Figure 9 indicated a better match of the dose distribution predicted by the DOSE-PYFER to the ground truth. Similarly, the visual results in Figure **10** confirmed that the dose distribution predicted by the proposed model is closer to the ground truth. The quantitative results of comparing the models are presented in **Table 6**. In the PTV region, the proposed model significantly decreases the disparity in PTV63 and PTV56 metrics compared to the C3D model by 0.53 $D_{95}$ and 0.44 $D_{99}$ in PTV56, and 0.044 $D_{99}$ in PTV56. Moreover, the average differences of $D_{0.1CC}$ in the brainstem and $D_{max}$ in the larynx were reduced by 0.03 and 0.49, respectively. Generally, the proposed model worked better on the OpenKBP dataset.

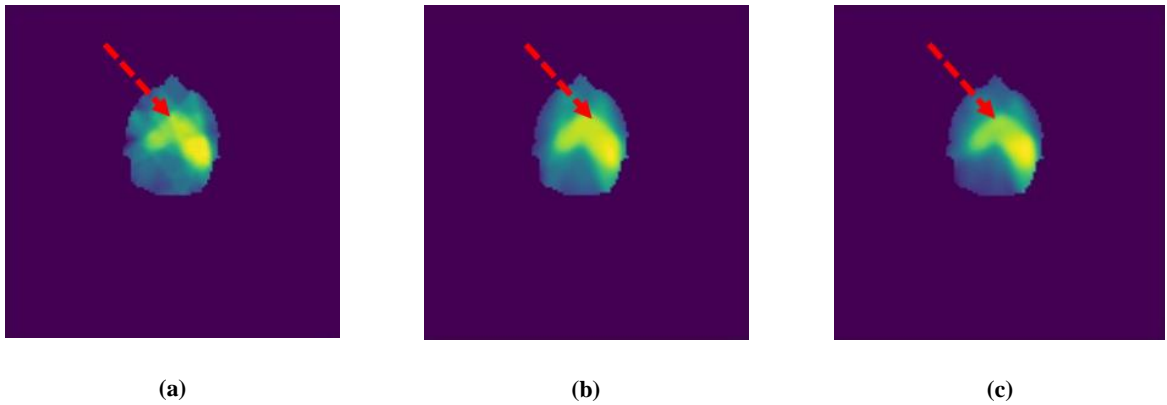

(a)           (b)           (c)

**Fig.7**. Comparison of (a) the ground-truth with the dose distribution of (b) the state-of-the-art model and (c) the proposed DOSE-PYFER.

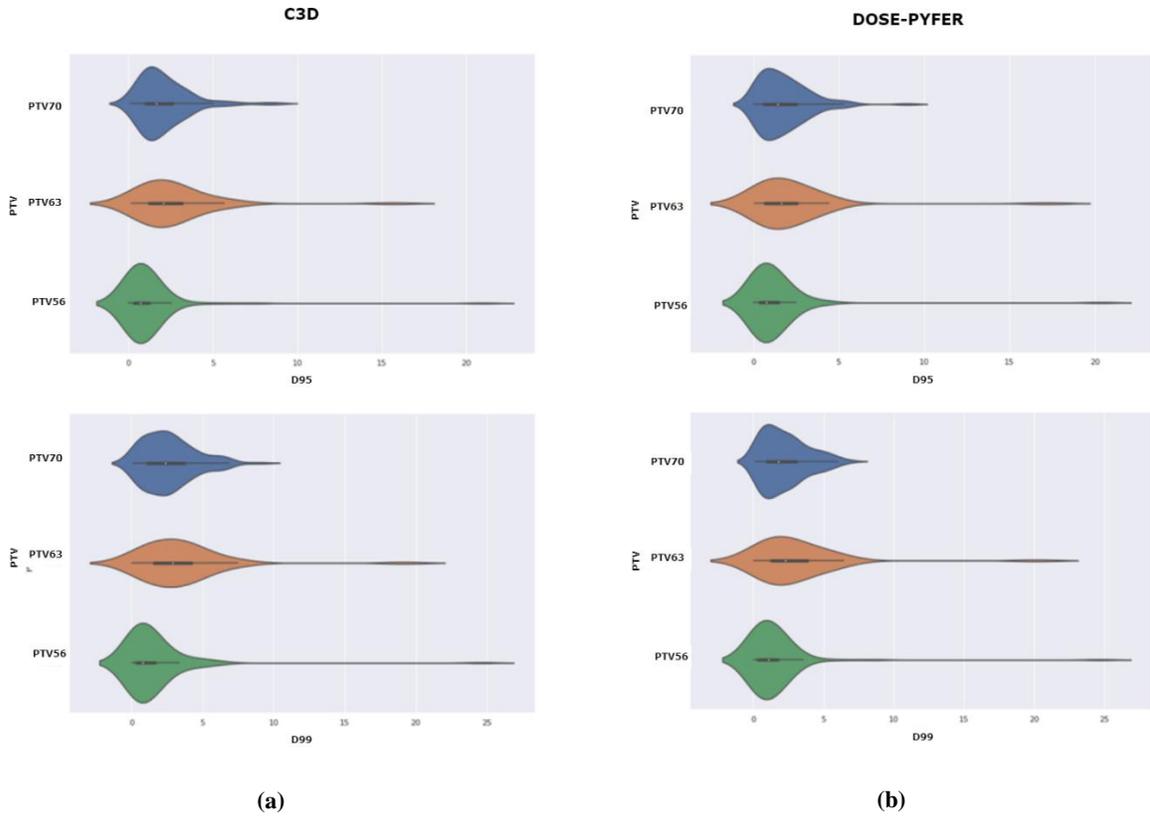

Fig.8. Comparison of violin plots for (a) the state-of-the-art model and (b) the proposed DOSE-PYFER.

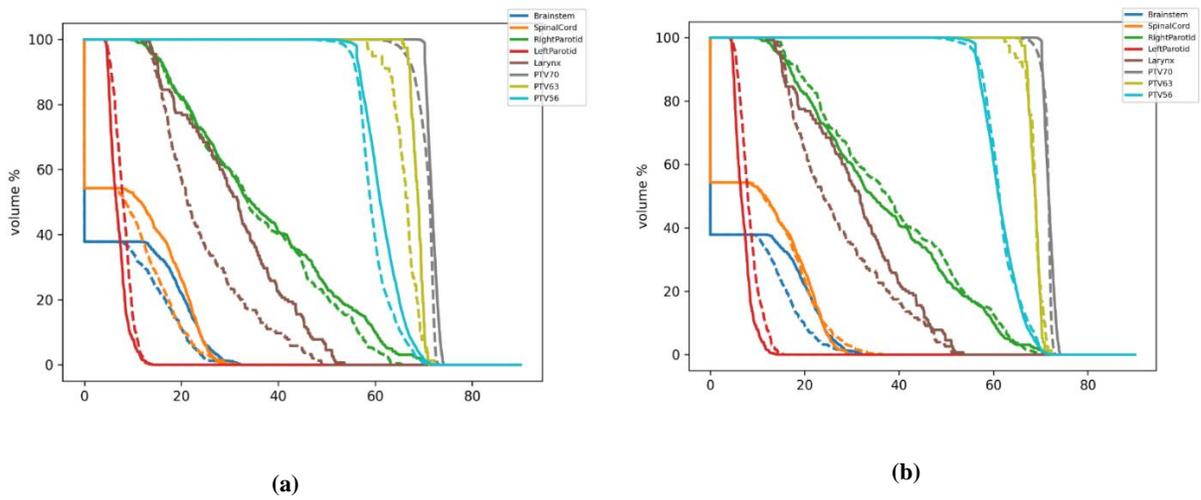

(a)

(b)

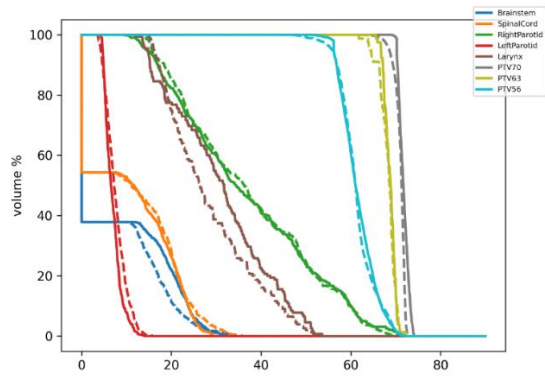

**(c)**

**Fig.9.** DVH comparison of (a) the proposed DOSE-PYFER, (b) the state-of-the-art model, and (c) the proposed end-to-end model with ground-truth.

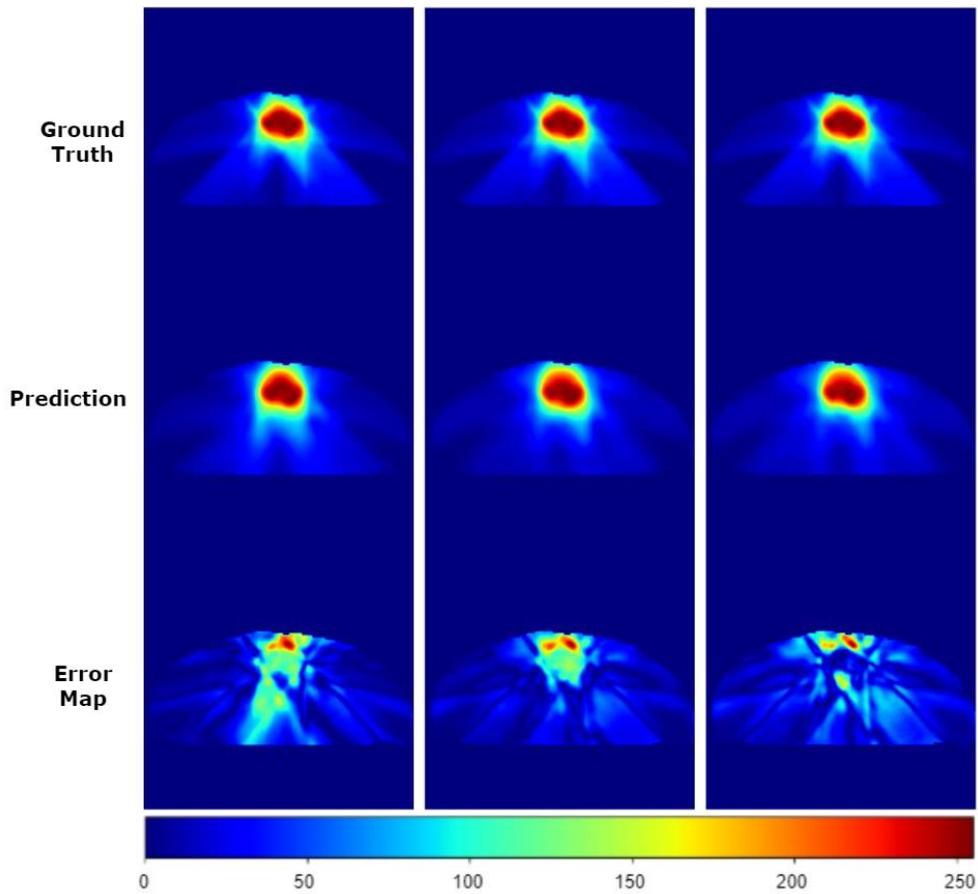

**Fig.10.** Qualitative comparison of the proposed DOSE-PYFER, the state-of-the-art model, and the proposed end-to-end model with ground truth, from left to right, respectively.

**Table 6**. Quantitative comparison of the proposed DOSE-PYFER with the state-of-the-art model and the proposed end-to-end model in terms of DVH metrics for PTV and OARs.

| ROI | | DVH Metrics | Method | | | | | | | | |
|---|---|---|---|---|---|---|---|---|---|---|---|
| | | | C3D Model | | | The proposed DOSE-PYFER | | | The proposed end-to-end model | | |
| | | | Result | Mean ± SD | P-value | Result | Mean ± SD | P-value | Result | Mean ± SD | P-value |
| PTV | PTV70 | D1 | 73.2770 | 1.5002±1.3670 | **0.0002** | 72.9166 | 1.6910±1.4664 | **0.0000<0.005** | 73.0964 | **1.3270±0.8865** | **0.0009** |
| | | D95 | 67.4298 | 2.0901±1.5634 | **0.0000<0.005** | 68.1440 | 1.7851±1.5064 | **0.0000<0.005** | 68.6431 | **1.3736±1.0865** | 0.0964 |
| | | D99 | 64.8316 | 2.6360±1.8437 | **0.0000<0.005** | 65.5527 | 2.2055±1.6088 | **0.0000<0.005** | 66.3525 | **1.4250±1.3299** | **0.0440** |
| | PTV63 | D1 | 71.3302 | 1.8922±1.4632 | **0.0051** | 71.4999 | 1.9008±1.4752 | **0.0229** | 71.6406 | **1.5385±1.1101** | 0.3420 |
| | | D95 | 61.4300 | 2.6868±2.5794 | **0.0058** | 62.6029 | **2.1574±2.6967** | 0.5028 | 62.4964 | 2.6370±3.9033 | 0.1724 |
| | | D99 | 59.0105 | 3.3991±3.1223 | **0.0046** | 59.5640 | **2.9552±3.2356** | **0.0365** | 59.4816 | 3.1477±4.9141 | 0.8318 |
| | PTV56 | D1 | 67.7707 | 1.7362±1.5981 | **0.0000<0.005** | 68.5741 | 1.5427±1.2366 | **0.0411** | 68.7403 | **1.0733±0.9639** | 0.8585 |
| | | D95 | 55.9416 | 1.1579±2.3387 | 0.8341 | 56.1814 | 1.2196±2.2201 | 0.2631 | 56.3652 | **1.0926±1.0378** | 0.4139 |
| | | D99 | 53.9360 | 1.5294±2.7901 | **0.0401** | 53.3245 | 1.4852±2.7577 | 0.8370 | 53.6977 | **1.1824±1.3636** | 0.9373 |
| OAR | Brainstem | $D_{0.1CC}$ | 32.8257 | 2.6152±3.0774 | **0.0418** | 31.8285 | 2.5852±3.1091 | 0.7601 | 32.3665 | **1.9156±1.8739** | 0.8832 |
| | | $D_{mean}$ | 8.9824 | 1.1795±2.0738 | 0.0512 | 8.6305 | 1.1351±2.1471 | 0.5931 | 8.7486 | **0.6557±0.7914** | 0.9867 |
| | Spinal Cord | $D_{0.1CC}$ | 34.7331 | 1.8017±1.6532 | 0.4151 | 35.6629 | 2.1073±1.7308 | **0.0126** | 35.6238 | **1.3976±1.3092** | 0.3255 |
| | | $D_{mean}$ | 16.7522 | 1.0161±0.9317 | 0.2619 | 17.1228 | 0.8816±0.7555 | 0.0966 | 15.8652 | **0.8804±0.6390** | **0.0200** |
| | Right Parotid | $D_{0.1CC}$ | 62.7437 | 1.9930±2.7804 | 0.5008 | 63.3151 | 1.9929±2.7531 | 0.3215 | 64.7608 | **1.5182±1.2271** | 0.2074 |
| | | $D_{mean}$ | 32.1555 | 1.3749±1.2070 | 0.4289 | 31.5994 | 1.6295±1.549 | 0.0687 | 32.2095 | **1.3282±1.4728** | 0.2189 |
| | Left Parotid | $D_{0.1CC}$ | 63.5126 | 1.7520±1.7106 | 0.1435 | 64.0342 | 1.7511±1.7029 | 0.5185 | 65.2137 | **1.4635±1.9530** | 0.6443 |
| | | $D_{mean}$ | 31.1606 | **1.2358±1.1073** | **0.0194** | 30.8818 | 1.5749±1.4566 | **0.0017** | 31.6466 | 1.3499±1.4932 | **0.0037** |
| | Esophagus | $D_{0.1CC}$ | 55.0113 | 2.9432±2.3369 | 0.5785 | 53.7021 | 2.8227±2.5369 | 0.0865 | 55.4704 | **1.6051±1.4552** | 0.5576 |
| | | $D_{mean}$ | 27.3970 | 2.0764±1.8287 | 0.4061 | 27.3519 | 2.0540±2.2053 | 0.3890 | 26.0315 | **1.5209±1.6697** | 0.0719 |
| | Larynx | $D_{0.1CC}$ | 66.2560 | 1.9459±1.5750 | 0.3440 | 66.6661 | 1.7805±1.4234 | 0.7939 | 67.8765 | **1.6872±1.3769** | 0.8820 |
| | | $D_{mean}$ | 46.8712 | 2.3134±2.1029 | **0.0113** | 47.7234 | 1.8269±1.9136 | 0.5605 | 50.6859 | **1.1697±1.1818** | 0.1173 |
| | Mandible | $D_{0.1CC}$ | 69.3696 | **1.7686±1.9316** | 0.5293 | 68.6174 | 1.9450±1.9993 | 0.0892 | 68.8418 | 1.8089±2.1393 | 0.9154 |
| | | $D_{mean}$ | 40.1036 | 1.7134±1.4318 | 0.9909 | 39.3458 | 1.8037±1.6951 | **0.0079** | 40.1813 | **1.5399±1.4261** | **0.0002** |
| Total | | Dose score | | 2.8568 | | | 2.7779 | | | **2.6429** | |
| | | DVH score | | 1.7544 | | | **1.7103** | | | 1.7523 | |

## 3.3 Effectiveness of the End-to-End Training of OAR-TRANSEG and DOSE-PYFER

We studied the potential contribution of the anatomical information derived from the segmentation task on the accuracy of dose prediction. In this way, the OAR-TRANSEG used a CT image as the input, and then its output with the contoured PTV was used as the input for the dose distribution prediction model to make an end-to-end model. The results showed that the dose score improved to 2.64 after end-to-end training. In Figure 9, comparing the DVH curves showed a better coincide between the solid lines to the corresponding dashed lines. Moreover, introducing the segmentation task clearly reduced the difference in Figure 10. For a precise evaluation, the DVH metrics for OARs and PTV were separately calculated in the end-to-end model. It is worth noting that most DVH metrics were improved by adding the OARs segmentation subnet (OAR-TRANSEG) to the proposed DOSE-PYFER subnet in end-to-end training. For instance, the end-to-end model significantly improved $D_{95}$ in PTV70 by 0.41 (23% of the proposed model) and 0.72 (34% of the C3D model). As shown in Table 6, the end-to-end model achieved overwhelming dose prediction results with the minimum average differences for PTV70, PTV56, and all OARs. The experimental results proved the effectiveness of end-to-end training of the OAR-TRANSEG and DOSE-PYFER.

*3.4 Other Ablation Studies*

*3.4.1 Cross-validation study*

We used a 6-fold cross-validation strategy to evaluate the generality of our model. Thus, 200 subjects were leveraged as the training set and 40 subjects as the validation set based on the split made by the OpenKBP dataset. To verify the generalization of our model, we computed and plotted the isodose volume similarity using the Dice parameter. The visualization results (Figure 11) illustrated a small distance between curves, indicating similarity between the curves of the validation and the test datasets. Also, this similarity was quantitatively verified as the average Dice was 0.94 and 0.93 for the validation and test datasets, indicating the generalization of our model.

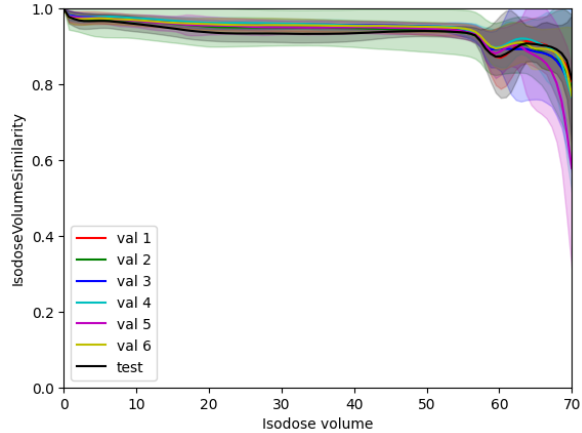

**Fig.11**. Comparison of isodose volume similarity between the predicted and the true dose maps. The shadow represents the standard deviation.

*3.4.2 Selection of the hyper-parameters*

Several experiments were conducted on the OpenKBP dataset using MLflow and Ray packages to determine the best hyper-parameters. The experiments involved testing various hyper-parameters such as adjusting the number of layers (num-layers) and the number of heads (num-heads) in the transformer-based encoder, selecting the activation function of Mish (Misra 2019) and ReLU (Agarap 2018), and modifying the values of $\lambda_2$ in the loss function, learning rate, and weight decay.

In the DOSE-PYFER architecture, features were extracted from the transformers in the transformer-based encoder. We selected four transformers in equally-spaced positions, where the output of the last transformer corresponded to the final output. Thus, possible values for the "num-layer" hyper-parameter were chosen as multiples of four based on this configuration. Also, the selection ranges for the hyper-parameter of num-heads in each transformer was the values of 3, 6, and 12. According to the dose score, the best values for num-layers and num-heads were 8 and 6, respectively. Additionally, we noticed that using the Mish activation function improved the dose score, whereas the ReLU activation function had the opposite effect. Moreover, we conducted experiments by varying the value of $\lambda_2$ from 1 to 9 since the value of $\lambda_1$ was fixed at 10. We found that the dose score improved as the value of $\lambda_2$ increased from 1 to 6, except for $\lambda_2=5$. However, when λ2 exceeded 8, the dose score dramatically decreased to the worst value. Hence, λ2 was set to 8. In the case of learning rate, and weight decay hyper-parameters, we selected the range of [10⁻

[10$^{-4}$, 10$^{-1}$] and [10$^{-4}$, 10$^{-3}$] for these parameters, respectively, based on the previous studies. Finally, the optimal values of learning rate and weight decay were set as 6.13×10$^{-4}$ and 1.63×10$^{-4}$, respectively. The quantitative results of selecting these hyper-parameters were given in Figure 12.

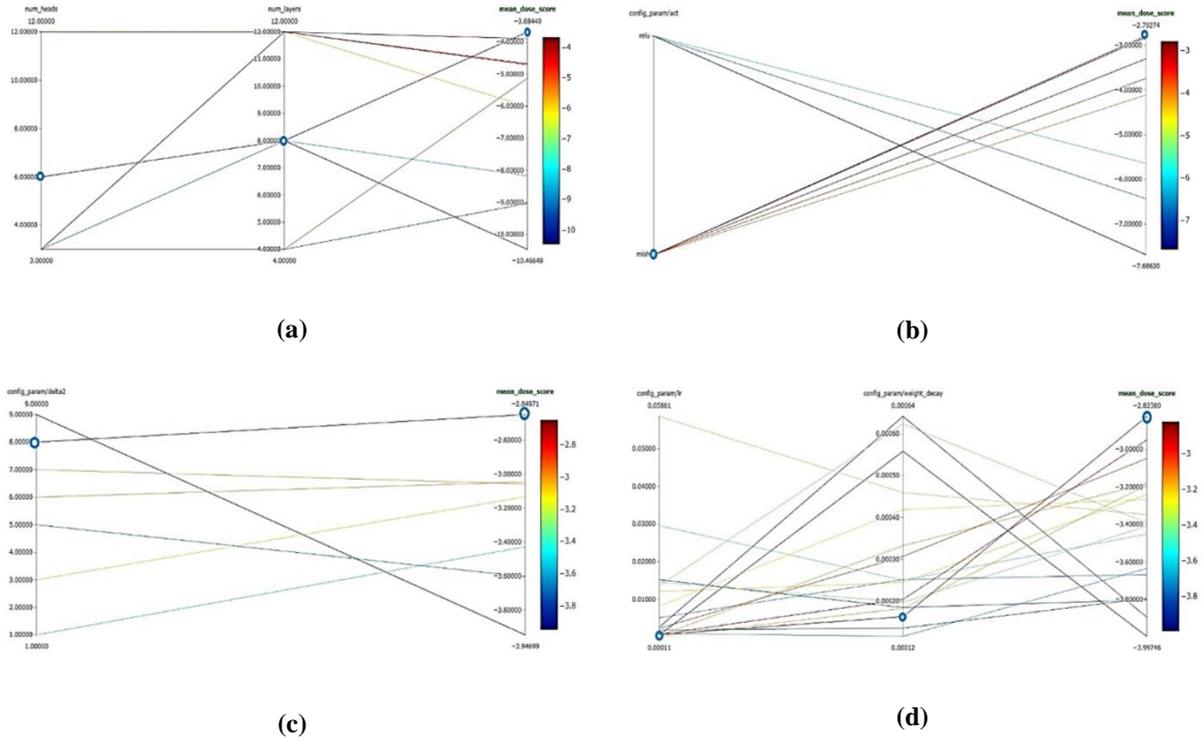

**Fig.12**. The quantitative impact of different hyper-parameters. (a) num-layers and num-heads, (b) ReLU and Mish activation functions, (c) λ$_2$ value, and (d) learning rate and weight decay.

## 4. Discussion

The study focused on developing a new deep learning-based model for automatic dose distribution prediction for H&N cancer radiotherapy. The previous works have some drawbacks. For instance, the anatomical information of OARs was neglected, which resulted in poor dose prediction accuracy. In this work, we proposed an encoder-decoder network called OAR-TRANSEG for OARs segmentation and a cascade encoder-decoder architecture called DOSE-PYFER for the dose distribution prediction. Finally, by combining OAR-TRANSEG and DOSE-PYFER subnets in end-to-end training, we proposed a model that uses the original CT images and the corresponding contoured PTVs to automatically delineate OARs and predict the dose distribution maps. We performed an extensive series of experiments for hyper-parameter tuning and selecting the best combination of them. Afterwards, the comparison tests were performed using the state-of-the-art models for each task to confirm the superiority and generality of the proposed model. The qualitative and quantitative superiority of both OARs segmentation and dose distribution prediction subnets confirmed for H&N cancers. We carried out ablation experiments to show the impact of different components on the proposed structures. Furthermore, we demonstrated that our model outperformed in the target regions with low prescribed doses, which is an important issue in radiation treatment.

In the proposed architecture, we used several strategies to extract the local and global features simultaneously. First, transformers were used as feature extraction tools for the global features in conjunction with the convolutional units. Secondly, we used multi-scale convolutional units. To improve the final output, we used the intermediate outputs of the network to calculate the loss function, whereas conventional methods increased the number of layers to achieve suitable performance. The intermediate outputs, as the scaled-down versions of the final output, were leveraged during the training to improve the final result. We also concluded that the main dose prediction task benefited from the anatomical features extracted by the auxiliary segmentation task, thereby improving its performance.

## 5. Conclusion

In this study, we proposed two novel architectures, including OAR-TRANSEG for OARs segmentation and DOSE-PYFER for dose distribution prediction in H&N cancer radiotherapy. The original CT images with the corresponding contoured PTVs are used to segment OARs and generate dose distribution maps. A proposed encoder-decoder architecture using transformers in the encoder and the multiscale convolutional units with skip connections in the decoder provides the conditions for extracting local and global features simultaneously and obtaining a better result. End-to-end training of the OAR-TRANSEG and DOSE-PYFER subnets made the process of predicting dose distribution maps based on IMRT treatment planning to be done automatically. In addition to increasing the accuracy of OARs segmentation and improving the performance of dose distribution prediction compared to previous studies, the proposed model improves the performance in regions with low prescribed doses. As further work, the automation of treatment planning for other tissues can be done. Also, developing bulk datasets is very important to achieve more accurate architectures to enable using these automatic techniques for clinical applications.

## Acknowledgement

The Authors wish to acknowledge Kermanshah University of Medical Sciences for their support, cooperation and assistance throughout the study (Grant no. 990688).